\documentclass[aps,prl,twocolumn,groupedaddress,showpacs]{revtex4}

\usepackage{graphicx}

\begin{document}


\title{Order and mobility of solid vortex matter in oscillatory driving currents}


\author{S. O. Valenzuela}
\altaffiliation[Present address: ]{Physics Department, Harvard
University, Cambridge, MA 02138.} 
\affiliation{Laboratorio de Bajas Temperaturas, Departamento de
F\'{\i}sica, Universidad Nacional de Buenos Aires, Pabell\'on I,
Ciudad Universitaria, 1428 Buenos Aires, Argentina}

\author{}
\affiliation{}


\date{\today}

\begin{abstract}

We study numerically the evolution of the degree order and
mobility of the vortex lattice under steady and oscillating
applied forces. We show that the oscillatory motion of vortices
can favor an ordered structure, even when the motion of the
vortices is plastic when the same force is applied in a constant
way. Our results relate the spatial order of the vortex lattice
with its mobility and they are in agreement with recent
experiments. We predict that, in oscillating applied forces, the
lattice orients with a principal axis perpendicular to the
direction of motion.

\end{abstract}


\maketitle


Two limiting cases have been distinguished in the de-pinning
process of a steady driven vortex lattice (VL) in type-II
superconductors. For a high density of weak short-range pinning
centers, the motion of the VL is inhomogeneous only in a narrow
region near the critical force, $F_c$ (which is determined by the
collective pinning theory \cite{larkin}), whereas for strong
pinning centers (or a small concentration of them) plastic
deformations become important and the motion is disordered
\cite{brass,koshelev}. In the last case, the VL orders and shows
elastic flow only for high enough driving forces
\cite{kv1,kv2,olson}(above a threshold force, $F_T$).

Recently, there has been much interest in the evolution of the
order and mobility of the VL for different thermomagnetic
histories and different temporal dependencies of the driving
force, $F^L$, which is provided by a flowing current. This is
motivated by the observation of a variety of puzzling phenomena
that include history-dependent dynamic response and memory effects
\cite{hend98,xiao,SOV01}. The nature of the ac response observed
in transport and ac susceptibility measurements has made clear
that it is not possible to directly extrapolate the
force-dependent evolution of a steady driven VL to explain the new
experimental findings for oscillatory forces. Some results are
intrinsic to the oscillatory dynamics. In particular, recent
experiments suggest that while for a given intensity of the dc
current the VL disorders, an ac current of the {\it same}
magnitude assists the VL in ordering \cite{hend98,SOV01}. This
conclusion, however, is indirectly drawn from changes in the VL
mobility. A key issue that naturally surges is the microscopic
dynamics of vortices and the relation between VL's order and
mobility with different temporal evolutions of the driving force.
Nevertheless, in spite of the growing amount of experimental
results on the ac dynamics of vortices, this relation has not been
investigated numerically up to now.

In this Letter, we present a numerical study that addresses
simultaneously the order and mobility of the VL in ac and dc
driving forces. We show that many of the experimental observations
can be reproduced for a VL around the crossover between plastic
and elastic regimes. For $F^L <F_c$, the VL is at rest whereas for
$F^L>F_T$ the vortex motion is ordered. For $F_c < F^L<F_T$, if
the force is applied steadily, the vortex motion is always plastic
and disordered but, if it is applied in an alternating way, it can
bring the VL to a more ordered state. We also find that a lower
density of defects in the VL is concomitant with a higher mobility
and that the VL tends to order with its principal lattice vector
aligned {\it perpendicular} to the direction of motion in contrast
with the orderly array of vortices in the steady force case
\cite{schmid}. All of these observations may have important
implications on the interpretation of recent neutron scattering
experiments \cite{ling}.

We consider a transverse 2D slice (in the $x-y$ plane) of an
infinite superconducting slab that contains rigid vortices that
are parallel to the sample surface
($\mathbf{H}$=$H\mathbf{\hat{z}}$). The overdamped equation of
motion of a vortex in position $ \mathbf{r}_i$ is $\mathbf{F}_i =
\sum_{j\neq i}^{N_v}\mathbf{F}^{vv}(\mathbf{r}_i-\mathbf{r}_j)+
\sum_k^{N_p}\mathbf{F}^{vp}(\mathbf{r}_i-\mathbf{r}_k^{p}) +
\mathbf{F}^L+\mathbf{F}^T_i= \eta\frac{d\mathbf{r}_i}{dt}$ where
$\mathbf{F}_i$ is the total force on vortex $i$ due to
vortex-vortex interactions ($\mathbf{F}^{vv}$), pinning centers
($\mathbf{F}^{vp}$), the driving current $\mathbf{J}$
($\mathbf{F}^{L}\sim\phi_0 \mathbf{J}\times \mathbf{\hat{z}}$) and
thermal fluctuations ($\mathbf{F}^T_i$). Here, $\eta$ is the
Bardeen-Stephen friction coefficient, $N_v$ the number of
vortices, $N_p$ the number of pinning sites and $\mathbf{r}_k^{p}$
the location of the $k$th pinning center. The vortex-vortex
interaction is modelled by a modified Bessel function
$\mathbf{F}^{vv}(\mathbf{r}_i-\mathbf{r}_j)=\frac{\phi_0^2}{8\pi^2
\lambda^3}$ $ A_vK_1$
$(|\mathbf{r}_i-\mathbf{r}_j|/\lambda)\mathbf{\hat{r}}_{ij}$ where
the dimensionless prefactor, $A_v$, tunes the rigidity of the VL
\cite{olson}. We cut off this interaction at $4\lambda$. Quenched
random disorder is modelled by pinning centers in uncorrelated
random positions that exert an attractive force on vortices:
$\mathbf{F}^{vp}(\mathbf{r}_i-\mathbf{r}_k^{p})= -A_p\:
e^{-\left(\frac{r_{ik}}{r_p}\right)^2}\mathbf{r}_{ik}$. $A_p$
tunes the strength of this force and $r_p$ its range. The thermal
force was implemented with a Box-Muller random number generator
and has the properties  $\langle F^T_{i\mu}\rangle=0$ and $\langle
F^T_{i\mu}(t)F^T_{j\nu}(t')\rangle=2 \eta k_BT
\delta_{ij}\delta_{\mu\nu}\delta(t-t')$ for a given temperature
$T$. We normalize length scales by $\lambda$ and forces by
$f_0=\frac{\phi_0^2}{8\pi^2 \lambda^3}$. We consider a sample with
periodic boundary conditions in the $x-y$ plane and size $L_x
\times L_y$ with $L_y/ L_x= \sqrt{3}/2$. The normalized vortex
density is $n_v=N_v \lambda^2 /L_x L_y=B \lambda^2/\phi_0$. We
choose $n_v = 1$ and a much higher density of pinning centers $n_p
= 25$ with $r_p=0.2\lambda$ so that the resulting random pinning
potential varies smoothly on a length scale of the order of, or
higher than, the coherence length, $\xi$ ($<\lambda/10)$. $A_p$
has a Gaussian distribution with central value $A_p^m=0.02f_0$ and
a standard deviation of 0.1$A_p^m$. System sizes from 400 to 1600
vortices have been investigated. We show results for 1600 vortices
and $4\times 10^4$ pinning sites and we set $T \sim 0$ (no thermal
noise) unless noted to the contrary.

History effects have been frequently associated with changes in
the density of defects in the VL. If this were the case,
information about the history of the sample could not be retained
either by a perfectly ordered VL or by a completely amorphous one,
i.e. it would be required a balance between the elastic and
pinning energies. In dynamical terms, this implies that, in
samples that present a path dependent response, the VL would be in
the region of the plastic to elastic crossover
\cite{brass,koshelev}. We start by identifying this crossover by
looking at the shape of the velocity-force dependence ($V-I$), the
vortex trajectories and their translational order for different
degrees of softness of the VL which are attained by changing $A_v$
\cite{olson}. Moving vortices induce a total electric field
$\mathbf{E}=\frac{B}{c}\mathbf{v}\times\mathbf{\hat{z}}$, with
$\mathbf{v}=\frac{1}{N_v} \sum_i \frac{d\mathbf{r}_i}{dt}$. To
quantify the degree of order of the VL we compute the structure
factor,
$S(\mathbf{k})=\frac{1}{N_v}|\sum_{i=1}^{N_v}e^{j\mathbf{k.r}_i}|$,
and determine the average concentration of vortices with
coordination number not equal to 6, $n_{def}$, using the Delaunay
triangulation procedure. As stated in Ref. \cite{koshelev}, in the
plastic regime there is a marked upward curvature in the $V-I$
curve which is not present in the elastic region. In addition, the
threshold force $F_T$ diminishes with $A_v$ and when approaching
to the elastic regime it mingles with $F_c$. This provides a
simple criterion to identify the crossover that, in our case,
takes place at $A_v \sim 0.85-0.9$. Following this result, we
perform the simulations at fixed $A_v=0.8$, which is in the region
of the crossover but in the plastic side \cite{koshelev,com1}.

In order to study the oscillatory dynamics and, at the same time,
to be able to directly compare the obtained results with those of
steady forces, we choose \textit{square} oscillating forces of
strength $F^L$ and frequency $\omega=2\pi/P$
($\mathbf{F}^L=F^L\mathbf{\hat{x}}$). For \textit{steady} driving
forces between $F_c \sim 0.0085 f_0$ and $F_T=0.012 f_0$ the
vortex flow is \textit{always} disordered with a high density of
defects, $n_{def}$ (of the order of 20\%). For higher driving
forces, $n_{def}$ diminishes and the movement is ordered with all
of the vortices moving at the same average velocity
\cite{koshelev,kv2}. Experimental results \cite{SOV01} suggest
that when vortices perform an oscillatory motion whose amplitude
is of the order of the lattice constant, the healing of defects
should be important. For this reason, we initially choose
$F^L=0.0118f_0$ and $P$ so that $F^L \: P/4\sim 1$ (in normalized
units).

\begin{figure}[t]
\includegraphics[width=3.3in]{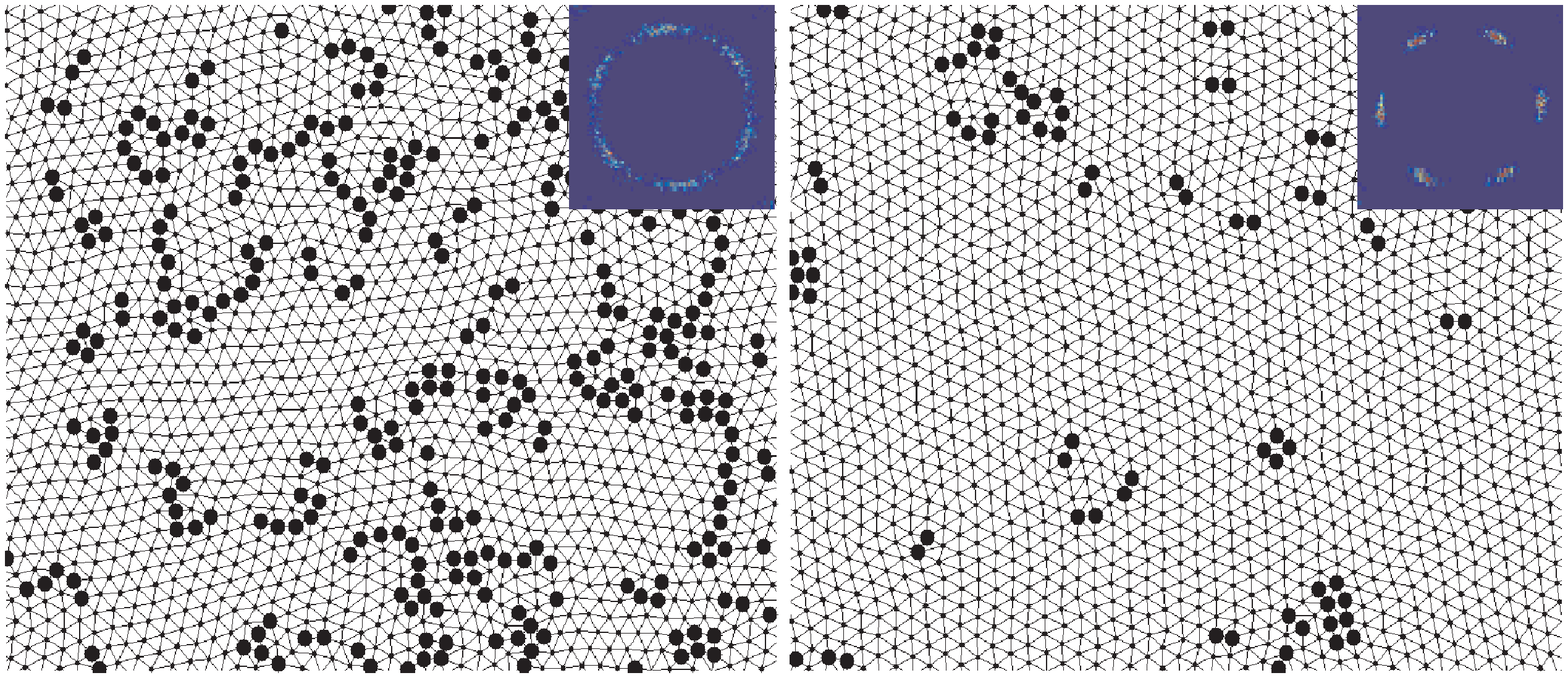} \vspace{2mm}
\caption{Delauney triangulation and structure factor,
$S(\mathbf{k})$ for steady forces (left) and oscillatory
square-forces (right). The more-ordered vortex configuration at
the right panel was obtained after 500 cycles of the oscillatory
force applied to the disordered state in the left panel. See text.
$A_v=0.8$ and $F_L=0.0118f_0$ (in the horizontal direction).}
\label{fig1}
\end{figure}

\begin{figure}[t]
\vspace{-9mm}
\includegraphics[width=2.7in]{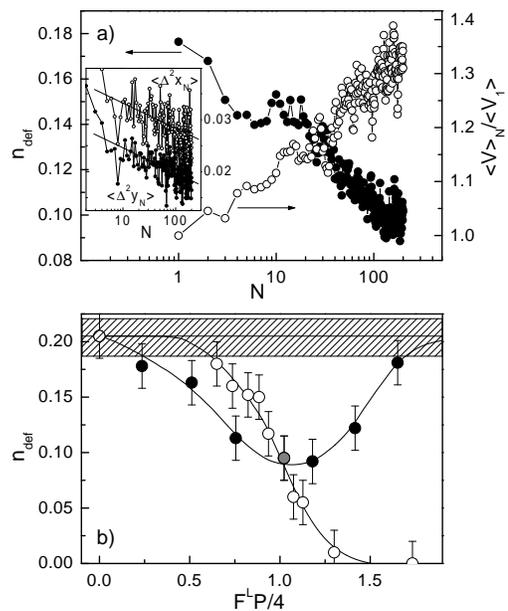} \vspace{-2mm}
\caption{Dynamical ordering. (a) Defect concentration, $n_{def}$,
average absolute-velocity, $\langle V \rangle$, and average
quadratic displacements of vortices per cycle,  $\langle \Delta ^2
x_N\rangle$ and $\langle \Delta ^2 y_N\rangle$ vs. the number of
cycles, $N$. (b) $n_{def}$ as a function of $F^LP/4$ after 100
cycles (filled circles: $F^L$ fixed, open circles: $P$ fixed). The
point $F^LP/4=1$ (gray) corresponds to $F^L=0.0118f_0$ and is
common to both curves. The shaded area marks the initial
disordered state. The lines are a guide to the eye.} \label{fig2}
\end{figure}

Some of the main results of our investigation are presented in
Fig. \ref{fig1} and \ref{fig2}. Fig. \ref{fig1} clearly pictures
the healing of defects that results from the oscillatory motion of
vortices. We start with a perfectly ordered VL. Then we apply a
steady force $F^L=0.0118f_0$ in the horizontal direction. The left
panel shows a snapshot of the configuration of the VL after
reaching a stationary state in which the average velocity of
vortices and the average concentration of defects remained
constant. This state presents a relatively high concentration of
defects. We then turn to an alternating driving force which
reverses periodically but keeps the same strength ($0.0118f_0$).
This alternating force is applied to the disordered state in the
left panel. After 500 cycles, it obtained the configuration of
the VL shown in the right panel. The corresponding structure
factors of both configurations are also shown. The Delaunay
triangulation and the structure factor reveal an important
reduction in $n_{def}$ when the oscillatory force is applied. It
is fundamental to emphasize that the initial state was obtained
with a steady force with the \textit{same strength} starting with
a \textit{perfectly ordered} VL. This implies that, for certain
values of the applied force, the flow of vortices can be plastic
and disordered introducing defects in the VL but, if the
\textit{same} force is applied in an alternating way, the defects
heal and the VL reorders.

In the Fig. \ref{fig2}a), we show the evolution of this
reordering. We plot the concentration of defects, $n_{def}$, the
average value of the absolute velocity, $\langle V \rangle$, and
the average quadratic displacements of vortices per cycle,
$\langle \Delta ^2 x_N\rangle$ and $\langle \Delta ^2 y_N\rangle$,
as a function of the number of cycles, $N$. We define $\langle
\Delta ^2 x_N\rangle =\frac{1}{N_v}
\sum_i^{N_v}[x_i(N)-\tilde{x}_i(N-1)]^2$ and $\langle \Delta ^2
y_N\rangle=\frac{1}{N_v} \sum_i^{N_v}[y_i(N)-\tilde{y}_i(N-1)]^2$,
where $\tilde{x}_i(N-1)=x_i(N-1)+\Delta X_{cm}(N)$ and
$\tilde{y}_i(N-1)=y_i(N-1)+\Delta Y_{cm}(N)$ with $(\Delta
X_{cm}(N), \Delta Y_{cm}(N))$ a correction due to the displacement
of the center of mass in the cycle $N$. The results shown
correspond to the average over 4 different random distributions of
the pinning centers. We see that, for an increasing number of
cycles, $n_{def}$ clearly decreases and the vortex mobility is
enhanced (as reflected by a 40\% increase in the average vortex
velocity). We also note that $\langle \Delta ^2 x_N,y_N\rangle$
diminish and that all magnitudes vary in an approximately
logarithmic way as observed experimentally \cite{hend98,SOV01}.
The combination of these observations indicates that, as the
number of defects decreases, vortices are clearly more mobile and
that, after completing one oscillation, they return closer to
their original positions ($\langle \Delta ^2 x_N\rangle$ and
$\langle \Delta ^2 y_N\rangle$ decrease). This suggests that
vortices organize and move more coherently as they are forced to
oscillate.

In Fig. \ref{fig2}b), we show what happens when we vary $P(F^L)$
while keeping $F^L(P)$ fixed. We plot $n_{def}$ as a function of
the parameter $F^LP/4$ after applying 100 cycles of the square
force to an initial disordered state as in Fig. \ref{fig1} and
Fig. \ref{fig2}a). An important result is that, if the period $P$
is increased so that the excursion of vortices greatly exceeds the
lattice constant, the VL distorts and does not reorder. This is
not surprising because if the amplitude of the oscillation is high
enough, the system should behave in the same way as when driven by
steady forces. This is a prediction that should be observed in
experiments as those in Ref. \onlinecite{SOV01}. For low enough
$P$, vortices oscillate in their pinning sites and the VL does not
reorder either.

Notably, if there is a tiny asymmetry in the amplitude of the
square force ($\sim 5\%$) the VL quickly disorders and, after a
few cycles, both $n_{def}$ and the vortex mobility reach values
close to those found with steady forces. This observation is
reminiscent to the experimental results reported in Ref.
\onlinecite{SOV01} where the vortex lattice is observed to quickly
become less-mobile when shaken by a sawtooth magnetic field. The
reduction in the mobility of the VL has been attributed to
disorder produced by a ratchet-like tearing of the VL due to the
different Lorentz forces involved in the ramp up and down of the
field \cite{SOV01}. We should note, however, that in our
simulations there is a net displacement of vortices after
completing one cycle of the asymmetric waveform that is not
expected to occur in the experiments.

An interesting issue surges in relation to the orientation of the VL.
At high-enough steady forces ($F^L >F_T$), a VL's principal vector
usually aligns with the direction of motion \cite{schmid}. Though
the details of this phenomenon are not completely understood,
Schmid and Hauger have argued that such an orientation is
preferred because it minimizes power dissipation \cite{schmid}.
They focused on the vortex motion at high velocities so that the
effect of the pinning potential can be handled as a small
perturbation.

\begin{figure}[t]
\includegraphics[width=2.5in]{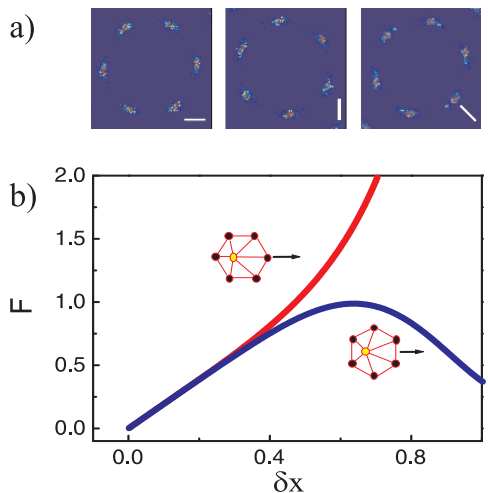} \vspace{-2mm}
\caption{(a) Structure factors of the vortex configuration after
``annealing" the sample from $T \sim T_m$ to $T=0$ with a square
force applied in the directions indicated by the corresponding
white line at the right-bottom corner of each image. It is seen
that the VL orients with one of its principal vectors
perpendicular to the direction of the force. (b) Restitutive force
considering one vortex moving $\delta x$ relative to its
neighbors. We consider two possible orientations of the VL.}
\label{fig3}
\end{figure}

It is less clear what would happen if the effect of pinning were
more important and the plastic distortions or the elastic energy
involved during the vortex motion increased. The perturbative
solution would fail to correctly describe the vortex behavior and
the elastic energy of the VL might play a preponderant role.
Indeed, in our simulations we find that, during the dynamical
reordering induced by symmetrical forces, a VL's principal vector
is aligned in most cases \textit{perpendicular} to the direction
of motion (see Fig. \ref{fig1}). In order to show that this
phenomenon is not due to the boundary conditions or to a
particular distribution of the randomly located pinning centers,
we performed a series of simulations where we slowly cooled the VL
from $T \simeq T_m$ \cite{com2} down to $T = 0$ while the VL was
subjected to the square force in different fixed directions. The
temperature was decreased in small steps after each cycle. After
600 cycles, $T = 0$. The resulting structure factor of the VL at
$T=0$ is shown in Fig. \ref{fig3}a) for three different directions
of the applied force (indicated by the white line). The
distribution of defects is the same for the three cases. The clear
tendency of the lattice to choose a transverse orientation (note
that the structure factor is rotated 90$^{\circ}$ in relation to
the real lattice) strongly suggests that this phenomenon is not
the result of either the boundary conditions or an orientation
favored by the particular distribution of defects. In fact, the
preferred orientation of the VL can be understood with a simple
energetic argument \cite{SOV01,com3}. In our calculations, the
vortex motion tends to be disordered and neighboring vortices move
at different average velocities. Some of the vortices can be
pinned and might not even move during a complete oscillation. The
oscillatory force produces a back-and-forth movement of vortices
and a consequent repeated interaction between neighbors. If the
overlap of vortices is avoided, the average elastic energy during
one period of the oscillation will be small. This favors an
ordered structure. Moreover, the most favorable condition will be
the one that minimizes the elastic energy. Because of that, we
believe that the elastic energy is a good quantity to estimate the
stability of the different orientations of the VL. To this end, we
consider one single vortex moving a distance $\delta x$ relative to its
neighbors, or vice versa, and we calculate the restitutive force
for different orientations [Fig. 3b)]. From this calculation, it
is straightforwardly seen that the lower elastic energy will
correspond to the lattice with the principal axis transversal to
the direction of motion, as observed in simulations. The same
conclusion is drawn if we consider the distortions that are
introduced in a nonrigid lattice.

Similar arguments have been used to explain the collective motion
of colloidal particles in oscillatory shear flow \cite{ackerson}.
The main and very important qualitative difference between the
colloidal and vortex systems is the presence of pinning centers in
the latter. However, the pinning centers favor an oscillatory
shear even when the applied force is uniform and, as a
consequence, the underlying mechanism of ordering is probably the
same in both systems.

In conclusion, we performed numerical simulations on the dynamics
of the vortex lattice that emphasize the intrinsic properties of
the oscillatory motion. We show that the vortex lattice may order
when forced to move with an oscillating force, even when the
motion that produces a steady force with the same absolute value
is plastic and disordered. As the vortex lattice orders, its
mobility increases and it moves in an increasingly coherent way.
The number of defects in the lattice and its mobility vary as the
logarithm of the number of cycles of the oscillating force. The
lattice orders because an ordered structure avoids the overlap of
neighboring vortices. For the same reason, the lattice orients
with one of its principal vectors perpendicular to the direction
of the force.

The author gratefully acknowledges discussions with V. Bekeris, L.
F. Cugliandolo, D. Dom{\'i}nguez, E. Y. Andrei, V. M. Vinokur, T. Giamarchi, M.
J. Rozenberg, G. Lozano, P. Tamborenea, M. J. S{\'a}nchez, F.
Laguna, V. Marconi, A. Kolton, and D. Elola.


\begin{thebibliography}
\bibitem{}
\bibitem{larkin} A.I. Larkin  and Y.N. Ovchinnikov, J. Low. Temp. Phys. {\bf 34}, 409 (1979).

\bibitem{brass} H. Jensen \textit{et al.}, \prl {\bf 60}, 1676 (1988); A. Brass, \textit{et al.}, \prb {\bf 39}, 102
(1989).

\bibitem{koshelev} A.E. Koshelev, Physica C {\bf 198}, 371
(1992).

\bibitem{kv1} P. Thorel \textit{et al.}, J. Phys. (Paris) {\bf 34},
447 (1973); S. Bhattacharya and M.J. Higgins, \prl {\bf 70}, 2617
(1993); U. Yaron \textit{et al.}, Nature {\bf 376}, 753 (1995);
M.C. Hellerqvist, \textit{et al.}, \prl{\bf 76}, 4022 (1996); W.
Henderson \textit{et al.}, \prl {\bf 77}, 2077 (1996); A. Duarte
\textit{et al.}, Phys. Rev. B {\bf 53}, 11336 (1996).

\bibitem{kv2} A.E. Koshelev and V.M. Vinokur, \prl {\bf 73}, 3580
(1994); M.C. Faleski \textit{et al.}, \prb {\bf 54}, 12427 (1996);
K. Moon \textit{et al.}, \prl {\bf 77}, 2778 (1996); S. Ryu
\textit{et al.}, \prl{\bf 77}, 5114 (1996); S. Scheidl and V.M.
Vinokur, \prb{\bf 57}, 13800 (1998). See also T. Giamarchi and P.
Le Doussal, \prl{\bf 76}, 3408 (1996); P. Le Doussal and T.
Giamarchi, \prb{\bf 57}, 11356 (1998); L. Balents \textit{et al.},
\prl{\bf 78}, 751 (1997) and \prb{\bf 57}, 7705 (1998); D.
Dom{\'{\i}}nguez, \prl{\bf 82}, 181 (1999); A.B. Kolton \textit{et
al.}, \prl{\bf 83}, 3061 (1999).

\bibitem{olson} C.J. Olson \textit{et al.}, \prl {\bf 81}, 3757
(1998).

\bibitem{hend98} W. Henderson \textit{et al.}, \prl {\bf 81}, 2352
(1998); E.Y. Andrei \textit{et al.}, J. Phys. IV {\bf 10}, 5
(1999); Y. Paltiel \textit{et al.}, Nature {\bf 403}, 398 (2000).

\bibitem{xiao} S.N. Gordeev \textit{et al.}, Nature {\bf 385},
324 (1997); Z.L. Xiao \textit{et al.}, \prl {\bf 83}, 1664 (1999);
S.O. Valenzuela and V. Bekeris, \prl {\bf 84}, 4200 (2000).

\bibitem{SOV01} S.O. Valenzuela and V. Bekeris, \prl {\bf 86}, 504 (2001); \prb {\bf 65}, 134513 (2002).

\bibitem{schmid} A. Schmid and W. Hauger, J. Low. Temp. Phys. {\bf 11}, 667
(1973). See also A.T. Fiory, \prl {\bf 27}, 501 (1971).

\bibitem{ling} X.S. Ling \textit{et al.}, \prl {\bf 86},
712 (2001); H. Safar  \textit{et al.} (unpublished).

\bibitem{com1} This value of $A_v$ is chosen to
keep the correlation length of the VL smaller than the size of the
system.

\bibitem{com2} Here $T_m$ is the melting temperature of the undisturbed VL.

\bibitem{com3} M.J. Rozenberg, private communication.

\bibitem{ackerson} B.J. Ackerson and P.N. Pusey, \prl {\bf 61}, 1033 (1989); W. Xue and G.S. Grest, \prl {\bf 64}, 419
(1990); H. Komatsugawa and S. Nos{\'e}, \pre {\bf 51}, 5944
(1995).

\end{thebibliography}
\end{document}